\documentclass[journal,comsoc]{IEEEtran}
\usepackage[T1]{fontenc}
\usepackage{amsmath,graphicx,cite}
\usepackage{balance}
\usepackage{amssymb}
\usepackage{algpseudocode}
\usepackage{algorithm}
\usepackage{amsthm}
\usepackage{mathtools}
\usepackage{algorithm}
\usepackage{bbm}
\usepackage{epstopdf}
\usepackage{xcolor}

\DeclareMathOperator*{\mini}{minimize}

\DeclareMathOperator*{\maxi}{maximize}
\DeclareMathOperator{\sbto}{subject \text{ } to}

\newtheorem{proposition}{Proposition}
\newtheorem{lemma}{Lemma}

\title{Hybrid Data-Sharing and Compression Strategy for Downlink Cloud Radio Access Network}

\author{Pratik~Patil,~Binbin~Dai, and~Wei~Yu,~\IEEEmembership{Fellow,~IEEE}
\thanks{The authors are with the Electrical and Computer Engineering Department, 
University of Toronto, Toronto, Ontario M5S 3G4, Canada.  
Emails: \{ppatil, bdai, weiyu\}@comm.utoronto.ca.
This work is supported by Huawei Technologies Canada Co. Ltd. and by Natural Science and Engineering Research Council (NSERC) of Canada. Part of this work has been presented in \cite{PatilYu_ITA14} at the Information Theory and Applications (ITA) Workshop, San Diego, CA, U.S.A., January 2014, and in \cite{PratikEUSIPCO} at the European Signal Processing Conference (EUSIPCO), Nice, France, September 2015.}}

\begin{document}

\maketitle

\begin{abstract}
This paper studies transmission strategies for the downlink of a cloud
radio access network, in which the base stations are connected to a centralized
cloud-computing based processor with digital fronthaul or backhaul links. We
provide a system-level performance comparison of two fundamentally different
strategies, namely the data-sharing strategy and the compression strategy, that
differ in the way the fronthaul/backhaul is utilized. It is observed that the
performance of both strategies depends crucially on the available
fronthaul or backhaul capacity. When the fronthaul/backhaul capacity is low, the
data-sharing strategy performs better, while under moderate-to-high fronthaul/backhaul capacity, the compression strategy is superior.  Using insights from such a comparison, we propose a novel hybrid strategy, combining the
data-sharing and compression strategies, that allows for better control over
the fronthaul/backhaul capacity utilization. An optimization framework for the
hybrid strategy is proposed. Numerical evidence demonstrates the performance
gain of the hybrid strategy.  
\end{abstract}


\section{Introduction}
\label{sec:intro}

This paper considers a promising future cellular architecture, the Cloud Radio Access Network
(C-RAN) \cite{CRAN_book}, in which the base stations (BSs) are connected to centralized cloud-computing based servers via high-speed but finite-capacity digital (wireline or wireless) links.  These digital links are
referred to either as backhaul, when they carry digital data, or as fronthaul, when they carry compressed analog signals. For convenience, this paper uses the term fronthaul and backhaul interchangeably.
One of the benefits of the C-RAN architecture is that it provides an
ability for flexible allocation of radio and computing resources
across all the BSs managed by the same central processor and a
cost-effective path for upgrading the existing wireless infrastructure
for mobile service delivery.
But more importantly, it facilitates coordinated and cooperative signal
processing across the multiple BSs connected to the same central
processor.

This paper studies the downlink transmission in a C-RAN setting. In
the downlink C-RAN, the user data originate from the centralized cloud server and are destined for
the mobile devices distributed throughout a geographical area, while
the BSs act as \emph{relays} between the user terminals and the cloud.
In this sense, the downlink C-RAN can be modeled as a broadcast-relay
channel. If the backhaul links between the cloud processor and the BSs
have infinite capacities, the information theoretical capacity
analysis for this setting is straightforward, as the downlink C-RAN
becomes a vector broadcast channel. But in the practical implementation of the C-RAN
architecture, the backhaul links have \emph{finite} capacities. In this more realistic case, both the theoretical analysis and the optimal practical system design become much more difficult.
This paper studies transmission strategies for the downlink C-RAN with
finite backhaul capacities.

There are two fundamentally different transmission strategies
for the downlink C-RAN, depending on whether the joint precoding
operation is performed at the central processor or at the individual
BSs. First, this paper asks the question of how the limited backhaul
capacities influence the achievable rates in each strategy, and
compares their system-level performance under practical network
settings. Second, this paper proposes a novel hybrid transmission
scheme, which allows for better utilization of the finite-capacity
backhaul, by combining these two strategies.

The interference mitigation capability of C-RAN stems from its ability
to jointly encode the user messages across multiple BSs.  One way to
enable such joint precoding is to simply share each user's messages
with multiple BSs over the backhaul links. This backhaul transmission
strategy, called the \emph{data-sharing} strategy in this paper, is
analogous to the decode-and-forward relaying strategy. 
As an alternative strategy, the joint precoding of user messages can
also be performed at the cloud server rather than at the individual
BSs. In fact, one of the original motivations for C-RAN is to
entirely shift the baseband processing from the BSs to the
central processor making BS units as simple as possible for easy
deployments, maintenance, and upgrades.
In this case, the precoded analog signals are compressed and forwarded to 
the corresponding BSs over the finite-capacity fronthaul/backhaul links
for direct transmission by the BS antennas.  This approach, called the
\emph{compression} strategy in this paper, is akin to the
compress-and-forward relaying strategy. 

One of the key questions in the implementation of C-RAN is whether or not the
functionalities of the BSs should be entirely moved to the central processor,
or whether there is some benefit to having a functional split between the
central processor and the BSs.  One of the main contributions of the paper is
to answer this question.

In the data-sharing strategy, the BSs receive clean copies of the user
messages.  Thus the precise beamformed signals can be formed at the BSs.
However, carrying raw user data multiple times to multiple BSs across the
entire network consumes large amount of backhaul. Therefore, practical
implementation of data-sharing strategy must involve finite cooperation
clusters, where each user selects a subset of cooperating BSs and only those
BSs in its cooperation cluster receive its message.  In the compression
strategy, since the beamformed signals are computed at the central processor,
the cooperation cluster size can, in principle, be large. But, the analog
beamformed signals need to be compressed in order to be transported to the
remote antenna heads under limited fronthaul. The compression process
introduces quantization noises that limit the system performance. We remark
here that the feasibility of the joint cooperative signal processing also
depends crucially on the availability of the channel state information (CSI) at
the BSs and the central processor. The requirement for CSI is proportional to
the BS cooperation cluster size.  The data-sharing strategy typically has
smaller cluster size, so it typically requires less CSI than the compression
strategy.

Individually, both the data-sharing and compression strategies have
been studied in the context of C-RAN. However, a fair system-level
comparison between the two strategies under practical network settings
has been a challenging task due to the numerical difficulty in solving the corresponding network optimization problems involving user scheduling, beamforming, power control, along with the
optimization of clusters for the data-sharing strategy and the
optimization of quantization noise levels for the compression strategy. This
paper tackles such a system-level performance evaluation and tries to find the
conditions under which one strategy outperforms the other.

The paper further demonstrates that in a practical C-RAN setting with finite
backhaul capacity, instead of individual data-sharing or compression
strategies, a hybrid strategy that combines the two can improve the overall
system performance. We propose an approach where the central processor directly
sends messages for some of the users to the BSs, along with the compressed
version of the precoded signals for the rest of the users. The intuition behind
such an approach is that it is beneficial, in terms of backhaul capacity
utilization, to send clean messages for strong users while compressing the rest
of the interference canceling signals. To quantify the benefit of this hybrid
strategy, this paper proposes an optimization methodology to select which users
to do direct data-sharing and which ones to compress.

\subsection{Contributions}

The overall contributions of this paper are as follows. First, the
paper provides a system-level performance comparison of the
data-sharing and compression strategies under finite backhaul capacity
and practical network settings by adopting a unified network-wide
utility maximization framework applicable to both strategies. The
unified optimization methodology is based on an equivalence between
the weighted sum rate maximization problem and the weighted minimization
of sum mean squared error (WMMSE) problem. As compared to most prior work, we take into account the loss due to practical modulation schemes in terms of gap to capacity for both strategies, and introduce a similar notion of gap to rate-distortion limit for the compression strategy to account for the
quantization losses due to non-ideal quantizers used in practice. A main
novelty of the first part of this paper is a joint optimization of the
beamformers and quantization noise levels for the compression strategy, based
on the equivalence between the weighted sum rate maximization and the WMMSE problems.
	
Moreover, this paper proposes a novel hybrid transmission strategy that
combines the data-sharing and compression strategies that allows for better
utilization of the limited backhaul capacity. An optimization framework to
quantify the performance gains due to the hybrid strategy is developed.
Specifically, we develop a unified optimization framework that jointly
optimizes the network-wide beamformers, user selection for either data-sharing
or compression, and the quantization noise levels for the compressed signals.
This framework generalizes the frameworks for both the data-sharing and
compression strategies.

\subsection{Related Work}

As pointed out earlier, information theoretically, the downlink of C-RAN is an instance of a broadcast-relay network, where the BSs can be considered as relays. The capacity of such network is unknown. A general coding strategy for the broadcast-relay network is proposed in \cite{kannan_raja_viswanath} based on a combination of Marton coding for the general broadcast channel \cite{marton_bc} and a coding scheme for deterministic linear relay networks \cite{Avestimehr2011QMF}. However, unlike in the uplink of the C-RAN, which is an instance of a multiple-access-relay channel, where compress-and-forward strategies such as quantize-map-forward scheme of \cite{Avestimehr2011QMF}, or more generally noisy network coding \cite{Lim2011NNC}, are known to be approximately optimal (in the sense of constant gap to the cutset outer bound), there are no approximation results known on the capacity region for the downlink C-RAN setup. The main difficulty lies in the need for careful coordination among the codewords for multiple user messages at the central processor. In the uplink, there is no such need as the central processor decodes all the compressed signals and the original user messages jointly. In the downlink, the central processor can induce coordination among different codewords and the relays potentially need to decode carefully chosen parts of the messages. Recently a new coding scheme that combines Marton coding for single-hop broadcast channels and partial decode-forward for relay channels, called distributed decode-forward (DDF), is proposed for broadcasting multiple messages over a general relay network in \cite{Lim2014DistributedDecodeForward} that can achieve the capacity of arbitrary broadcast relay networks to within constant gap that only depends on the number of nodes in the system. In the context of C-RAN, the recent work \cite{PPWY2018_ddf_vs_compression} shows that under certain conditions the DDF strategy applied to C-RAN is equivalent to a generalized compression strategy that can achieve the capacity of downlink C-RAN to within a constant gap that only depends on the number of users and BSs. However, these results are information theoretic in nature, and a practical implementation of the generalized compression strategy is yet to be realized.

If the backhaul capacity is infinite, downlink C-RAN with a Gaussian channel model reduces to the well-known vector Gaussian broadcast channel, for which dirty paper coding (DPC) achieves the capacity region. For the finite backhaul capacity, however, DPC and other linear precoding schemes cannot be applied directly. In \cite{MarschFettweis09}, inner bounds for the downlink transmission schemes with different levels of BS cooperation (infinite, limited or no BS cooperation) are studied. The effect of imperfect CSI at the BSs and users is also taken into account.

For compression based strategies, a compressed version of DPC (named CDPC)
is introduced in \cite{SimeoneSomekhPoorShamai09}. Different
transmission strategies that require varying degrees of codebook
information (the encoding function information needed to employ DPC)
at the BSs are investigated for a simple Wyner type model. CDPC is 
a scheme in which the BSs are oblivious of any codebook information, 
and the central processor performs joint DPC, then independently
compresses the codeword for each BS, and sends the quantized codeword
to the corresponding BS.
If some degree of codebook information is available at the BSs, then
data-sharing becomes possible. The conclusion of
\cite{SimeoneSomekhPoorShamai09} is that oblivious BSs are sufficient
in the regime of sufficiently large backhaul capacity for the Wyner
model. To further optimize the compression strategy,
\cite{ParkSimeoneSahinShamai13} proposed a multivariate compression
scheme across the signals of all the BSs, instead of independent
compression for each BS, to better control the
effect of resulting total quantization noises at the users by
correlating the quantization noises for signals of different BSs. An
iterative algorithm achieving a stationary point for the problem of
maximizing sum rate with respect to the precoding matrix and the
quantization noise covariance matrix is proposed. Our work differs
from \cite{ParkSimeoneSahinShamai13} in that
\cite{ParkSimeoneSahinShamai13} optimizes the covariance matrices of
transmit beamformers along with the quantization noise covariance
matrix using a rank approximation method. In our optimization framework for
the compression strategy, we make a novel use of the equivalence
between the weighted sum rate maximization and the WMMSE problem, which does
not require any approximation.

For data-sharing based strategies, various ways to selectively share
the user messages have been investigated in the literature
\cite{NgEvansHanlyAktas08, ZakhourGesbert11,
ZhaoQuekLei13,Zhuang14,Zhang2014green}.
Information theoretic results for the downlink network
multiple-input multiple-output (MIMO) model
using the data-sharing strategy have been reported in
\cite{SimeoneSomekhPoorShamai09,MarschFettweis08,jing2008multicell},
but most of the theoretical works are limited to
certain simplified models. A modified linear Wyner cellular model is
studied in \cite{SimeoneSomekhPoorShamai09}, and a two-BS, two-user
setup is studied in \cite{MarschFettweis08}. Our optimization
framework for the data-sharing strategy is based on previous work on
sparse beamforming in \cite{DaiYu_Access14}, but extending the
algorithm in \cite{DaiYu_Access14} to account for the gap to capacity factor. Differing from 
\cite{ZhaoQuekLei13,Zhuang14,Zhang2014green}, this paper takes a
network utility maximization approach in order to provide a realistic
and unified comparison to the data-sharing strategy in terms of the 
cumulative distribution of user rates that accounts for fairness.

We mention here that the performance comparison conducted in this paper
is closely related to the work of \cite{KangSimeone16}, where a sum-rate 
evaluation has been carried out both under perfect and stochastic CSI.  
The results of this paper go one step further in examining the
cumulative distribution of user rates.  Further, we utilize a WMMSE
approach, instead of the stochastic successive upper bound
minimization method of \cite{KangSimeone16}, but our findings are
largely consistent with that of \cite{KangSimeone16}. 

It is worth pointing out that a third transmission strategy, based on the compute-and-forward (CoF) strategy for relay networks \cite{nazer_gastpar}, is proposed in \cite{hong_caire_journal} and named as reverse compute-and-forward (RCoF). The roles of BSs and users are reversed in RCoF compared to CoF. Since users do not cooperate, an appropriate invertible precoding is performed to the messages to be sent by the BSs at the central processor such that the effect of linear combination can be undone at the user terminal and each user obtains just its desired message in the end. Low complexity version of RCoF based on the standard scalar quantization is also studied, named as quantized reverse compute-and-forward.
But, as with CoF, the performance of RCoF is quite sensitive to the channel coefficients due to the non-integer penalty, since the channel coefficients are not exactly matched to the computed integer linear combination.
The main challenge with the lattice-coding based strategies is that the underlying optimization problems often involve integer matrices, which are difficult to optimize in practical networks.

\subsection{Organization}

The rest of the paper is organized as follows. Section \ref{sec:model}
describes the C-RAN system model under consideration. Sections
\ref{sec:data_sharing} and \ref{sec:compression} formulate the
network-wide optimization for the data-sharing and compression
strategies. The optimization frameworks for the two strategies are
provided in separate sections, then the numerical system-level
performance comparison between the two is made in Section
\ref{sec:data_sharing_vs_compression}. Section \ref{sec:hybrid}
proposes the hybrid strategy that combines the data-sharing and
compression strategies. We provide a unifying optimization framework
for the hybrid strategy that generalizes the data-sharing and
compression strategies. Joint optimization of network-wide
beamformers, user selection for data-sharing component, and
quantization noise optimization for the compressed signal is
performed. We then provide system-level numerical evaluation of the
hybrid strategy in Section \ref{sec:hybrid_evaluation} to quantify its performance gains over the individual
data-sharing and compression strategies. Finally, Section \ref{sec:conclusions} concludes the paper.

\subsection{Notation}

The notations used in this paper are as follows. Plain lower or upper case letters are used to denote scalars, e.g., $w$, $C$. Bold face lower letters are used to denote vectors, e.g., $\mathbf{w}$. Bold face upper letters are used to denote matrices, e.g., $\mathbf{H}$. An $n$-dimensional identity matrix is denoted by either $\mathbf{I}_{n \times n}$ or $\mathbf{I}$ when the dimension is clear from the context. For a complex scalar, $\text{Re}\{\cdot\}$ denotes its real part and $\vert \cdot \vert$ denotes its magnitude. For a vector, $(\cdot)^T$ denotes its transpose, $\vert|\cdot\vert|_p$ denotes its $\ell_p$ norm. For a matrix, $(\cdot)^{-1}$ denotes its inverse, $(\cdot)^H$ denotes its conjugate transpose (or just conjugate in case of a scalar). For a random variable, $\mathbb{E} \left [ \cdot \right]$ denotes its expected value. Calligraphy letters are used to denote sets, e.g., $\mathcal{L}$. Letters $\mathbb{C}$ and $\mathbb{R}$ are used to denote the set of real and complex numbers, respectively.

\section{System Model}
\label{sec:model}

Consider the downlink of a C-RAN comprising of $L$ BSs equipped with $M$ antennas serving $K$ users equipped with $N$ antennas. All the BSs are connected to a central processor with capacity-limited backhaul links or fronthaul links.
The capacity of the backhaul link connecting BS $l$ to the central processor is denoted by $C_l$, $l \in \mathcal{L} = \{1,\ldots,L\}$. We transmit a single independent data stream from the central processor to each user. The user $k$'s information signal is denoted by $s_k$, $k \in \mathcal{K} = \{1,\ldots,K\}$, and it is assumed to be chosen independently from a complex Gaussian distribution with zero-mean and unit variance. We assume that the central processor has access to the data and perfect CSI for all the users in the network. The complex signal transmitted by antenna $m$ at BS $l$ is denoted by $x_l^m$, $m \in \mathcal{M}=\{1,\ldots,M\}$, $l \in \mathcal{L}$. We assume a per-antenna transmit power constraint with maximum power budget denoted by $P_l^m$, i.e.,
\begin{equation}
\mathbb{E}[\left \vert x_l^m \right \vert^2] \le P_l^m, \quad l \in \mathcal{L}, m \in \mathcal{M}.
\end{equation}

A flat-fading channel model is assumed. Let $\mathbf{x}_l \in \mathbb{C}^{M \times 1} = [x_l^1,\ldots,x_l^m]^T$ denote the vector signal transmitted by BS $l$ and $\mathbf{x} \in \mathbb{C}^{LM \times 1}=[\mathbf{x}_1^T,\ldots,\mathbf{x}_L^T]^T$ denote the aggregate signal from all the BSs. The
received signal at user $k$, $\mathbf{y}_k \in \mathbb{C}^{N \times 1}$, is
\begin{equation}
\mathbf{y}_k=\mathbf{H}_k\mathbf{x}+\mathbf{z}_k,
\end{equation}
where $\mathbf{H}_k \in \mathbb{C}^{N \times LM}=[\mathbf{H}_{1,k},\ldots,\mathbf{H}_{L,k}]$ is the channel to user $k$ from all the BSs, $\mathbf{H}_{l,k} \in \mathbb{C}^{N \times M}$ is the channel response 
from the $M$ transmit antennas of BS $l$ to the $N$ receive antennas of user $k$, and $\mathbf{z}_k$ is the additive complex Gaussian noise with zero-mean and variance $\sigma^2$ on all of its diagonals.

\section{Data-sharing Strategy}
\label{sec:data_sharing}

\begin{figure}[t]
	\centering
	\includegraphics[width=1.0\columnwidth]{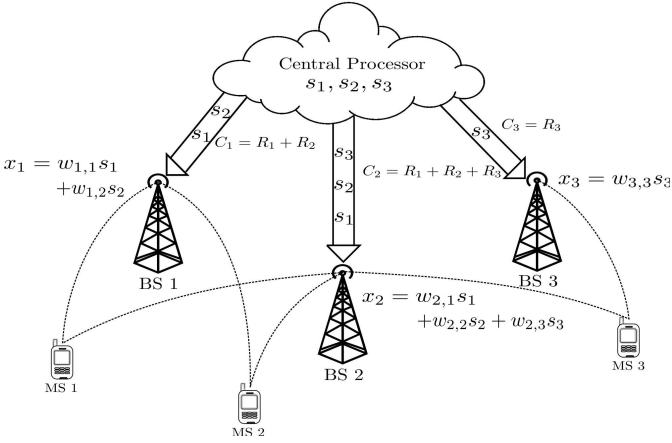}
	\caption{The data-sharing strategy for the downlink C-RAN.}
	\label{fig:figureDataSharing}
\end{figure}

In the data-sharing strategy, as shown in Fig.~\ref{fig:figureDataSharing}, a cluster of BSs locally form beamformers to cooperatively serve each user. The data for that user is replicated at all participating BSs in the cluster via the backhaul links. A crucial decision is to select an appropriate cluster of BSs for each user for interference mitigation, while being constrained under the limited backhaul capacity.

\subsection{Optimization Framework}

Let $\mathbf{w}_{l,k} \in \mathbb{C}^{M \times 1} =
[w_{l,k}^1,\ldots,w_{l,k}^M]^T$ be the beamforming vector from BS $l$ to user
$k$ with $w_{l,k}^m$ denoting the beamforming coefficient from the $m$th antenna of
BS $l$ to user $k$, and $\mathbf{w}_k \in \mathbb{C}^{LM \times 1} =
[\mathbf{w}_{1,k}^T,\ldots,\mathbf{w}_{L,k}^T]^T$ be the aggregate network-wide
beamformer to user $k$ from all the BSs. If user $k$ is not cooperatively
served by BS $l$, then $\mathbf{w}_{l,k} = \mathbf{0}$. This can be
equivalently represented by: $\left\|\mathbf{w}_{l,k}\right\|_2^{2} = 0$, if BS
$l$ does not serve user $k$.
The beamformed signal $\mathbf{x}$ to be transmitted by all the BSs can be written as
\begin{equation}
\mathbf{x}=\sum_{k=1}^K \mathbf{w}_k s_k.
\label{eq:beamformedSignal}
\end{equation}
At user $k$, the signal-to-interference-plus-noise ratio (SINR) is
\begin{equation}
\label{SINR_k}
\text{SINR}_k =
\mathbf{w}_{k}^{H} \mathbf{H}_{k}^{H} \left (\sum_{j \neq k } \mathbf{H}_{k} \mathbf{w}_j \mathbf{w}_{j}^{H} \mathbf{H}_{k}^{H} + \sigma^2 \mathbf{I} \right )^{-1} \mathbf{H}_{k} \mathbf{w}_k.
\end{equation}
The information theoretical achievable rate for user $k$ is related to $\text{SINR}$ as $R_k = \log(1+\text{SINR}_k)$.
However, this rate expression assumes Gaussian signaling, while in practice Quadrature Amplitude Modulation (QAM) constellations are typically used for the Gaussian channel in the moderate and high SINR regime.
To achieve a given data rate at a certain probability of error, we need an SINR higher than what is suggested above. This extra amount of power is usually captured by a so-called SNR gap, denoted here by $\Gamma_m$. Its value is approximately independent of the size of the constellation for square QAM, and can be easily computed as a function of the target probability of error~\cite{Forney1998modulation}. 
For example, at $P_e = 10^{-6}$, uncoded QAM has $\Gamma_m = 9$ dB. The use of error correcting codes can lower the value of $\Gamma_m$. Now with the SNR gap taking into account, we can rewrite the achievable rate for user $k$ as
\begin{equation}
\label{eq:Rk_DataSharing_MIMO}
R_k = \log \left( 1 +\frac{\text{SINR}_k}{\Gamma_m} \right).
\end{equation}
The optimization problem of finding the optimal set of BS clusters and
beamformers for the data-sharing scheme can now be formulated as a weighted sum
rate maximization problem under per-antenna power constraints and per-BS backhaul constraints as follows:
\begin{subequations} \label{WSRwithBkhaul_DataSharing_MIMO}
	\begin{align}
	\maxi_{\{\mathbf{w}_{l,k}\}} \quad & \sum_{k=1}^K \alpha_k R_k \\
	\sbto \quad &  \sum_{k=1}^K \left \vert w_{l,k}^m\right \vert^{2} \leq P_l^m, \quad l \in \mathcal{L}, m \in \mathcal{M} \label{powerconst_DataSharing_MIMO} \\
	& \sum_{k=1}^K \mathbbm{1}\left\{ \left\|\mathbf{w}_{l,k}\right\|_2^{2} \right\} R_k \leq C_l, \quad l \in \mathcal{L},  \label{bkhaulconst_DataSharing_MIMO}
	\end{align}
\end{subequations}
where $\alpha_k$ denotes the priority weight associated with user $k$ at the current user scheduling time slot which can be updated according to proportional fairness criterion, for example. The indicator function $\mathbbm{1}\left\{ \left\|\mathbf{w}_{l,k} \right\|_2^2\right\}$ in the constraint~(\ref{bkhaulconst_DataSharing_MIMO}) denotes whether BS $l$ participates in beamforming to user $k$, and if so, the user rate $R_k$ is included in the backhaul constraint $C_l$. The constraint~(\ref{powerconst_DataSharing_MIMO}) accounts for the per-antenna power constraint at antenna $m$ of BS $l$. The beamforming coefficients are computed at the central processor, and are assumed to be transmitted to the BSs without any error. We neglect the backhaul consumption for transmitting the beamforming coefficients, because the beamformers need to be transmitted only once during each user scheduling time slot; comparing with the backhaul needed to send the data, it is a very small fraction. The above formulation considers joint design of BS clustering, beamforming, and power control. Note that it also implicitly does joint user scheduling, which can be seen from the fact that a user $k$ is scheduled, i.e., $R_k$ is non-zero, if and only if its beamformer vector $\mathbf{w}_k$ is non-zero. Thus the user scheduling is implicitly jointly done along with BS clustering and beamforming optimization to satisfy the per-antenna and per-BS backhaul constraints. The optimization problem is solved repeatedly and the BS clusters are dynamically optimized in each time slot as the priority weights are updated.

\subsection{Optimization Methodology}

The presence of the backhaul constraint~(\ref{bkhaulconst_DataSharing_MIMO}) makes the optimization problem challenging. In this paper, we follow the approximation suggested in \cite{DaiYu_Access14} to first write the indicator function as a $\ell_0$ norm which is then approximated as a weighted $\ell_1$ norm as
\begin{equation}
\mathbbm{1}\left\{ \left\|\mathbf{w}_{l,k}\right\|_2^2 \right\} = \left\| \left\|\mathbf{w}_{l,k}\right\|_2^2 \right\|_0 \approx \beta_{l,k}\left\|\mathbf{w}_{l,k}\right\|_2^2,
\end{equation}
where $\beta_{l,k}$ is a constant weight associated with BS $l$ and user $k$ and is updated iteratively according to
\begin{equation}
\label{UpdateBeta_MIMO}
\beta_{l,k} = \frac{1}{\left\|\mathbf{w}_{l,k}\right\|_2^2 + \tau},
\end{equation}
for some regularization constant $\tau > 0$ and $\left\|\mathbf{w}_{l,k}\right\|_2^2$ from the previous iteration. This simplifies the constraint~(\ref{bkhaulconst_DataSharing_MIMO}) to
\begin{equation}
\label{approxbkhaulconst_DataSharing_MIMO}
\sum_{k=1}^K  \beta_{l,k} \left\|\mathbf{w}_{l,k}\right\|_2^2 R_k \leq C_l, \quad l \in \mathcal{L},
\end{equation}
which is equivalent to a generalized power constraint, if $R_k$ is assumed fixed and heuristically chosen from the previous iteration in an iterative manner. The resulting optimization problem then becomes:
\begin{subequations} \label{WSRwithBkhaul_approx_MIMO}
	\begin{align}
	\maxi_{\{\mathbf{w}_{l,k}\}} \quad & \sum_{k=1}^K \alpha_k R_k \\
	\sbto \quad &  \sum_{k=1}^K \left \vert w_{l,k}^m \right\vert^{2} \leq P_l^m, \quad l \in \mathcal{L}, m \in \mathcal{M}  \label{WSRwithBkhaul_approx_MIMO_PowerConstraint} \\
	& \sum_{k=1}^K  \beta_{l,k}\hat{R}_k \left\|\mathbf{w}_{l,k}\right\|^2 \leq C_l, \quad l \in \mathcal{L}, \label{WSRwithBkhaul_approx_MIMO_BackhaulConstraint}
	\end{align}
\end{subequations}
where $\hat{R}_k$ is the rate from the previous iteration.

Even though the approximated problem~(\ref{WSRwithBkhaul_approx_MIMO}) is still non-convex, it can be formulated as an equivalent WMMSE problem using the equivalence between the weighted sum rate maximization and the WMMSE problem.  The advantage of working with the WMMSE problem is that the optimization variables can be split into groups such that with respect to each group of variables, the optimization problem is convex, if all other variables are fixed.  Thus we can use the block coordinate descent method to reach a stationary point of~(\ref{WSRwithBkhaul_approx_MIMO}). The relationship between the weighted sum rate maximization and the WMMSE problem is first established for the MIMO broadcast channel in \cite{Cioffi08}, and is generalized to the MIMO interference channel in \cite{Tom_SP11} and to the MIMO interference channel with partial cooperation in \cite{KavianiSimeoneKrzymienShamai}. In the context of C-RAN, the equivalence is used in~\cite{DaiYu_Access14}. The difference between the formulation~(\ref{WSRwithBkhaul_DataSharing_MIMO}) and that in~\cite{DaiYu_Access14} is the gap factor $\Gamma_m$ and per-antenna power constraints, instead of per-BS power constraint. It can be verified that the equivalence between weighted sum rate maximization and WMMSE extends even for~(\ref{WSRwithBkhaul_approx_MIMO}).

We summarize the overall algorithm for the optimization of the data-sharing strategy in Algorithm~\ref{alg:LCSparse_MIMO}. The algorithm is essentially the one already proposed in \cite{DaiYu_Access14}. We include it here for completeness and for subsequent unified comparison to the compression strategy. Although we do not have theoretical guarantee of its convergence in general, it is observed to converge in simulations.
Following a similar analysis as in \cite{DaiYu_Access14}, we observe that finding the transmit beamformers by solving the optimization problem (\ref{QCQP_DataSharing_MIMO}) in Step 3 of the Algorithm~\ref{alg:LCSparse_MIMO} is computationally the most expensive step and dominates rest of the steps for any iteration. 
Problem (\ref{QCQP_DataSharing_MIMO}) is a 
quadratically constrained quadratic programming (QCQP) problem which can also be equivalently cast as a second order cone programming (SOCP) problem, and can be solved in $O((KLM)^{3.5})$ time complexity using the interior-point method \cite{Ye_InteriorPointAlgorithms_97} in standard solvers such as CVX \cite{cvx}. Thus, the overall computational complexity of Algorithm~\ref{alg:LCSparse_MIMO} scales as $O((KLM)^{3.5}T)$, where $T$ is the number of iterations to converge.

\begin{algorithm}[t]
	{\bf Initialization}: $\{\beta_{l,k}\}, \{\mathbf{w}_k\}, \{\hat{R}_k\} $; \\
	{\bf Repeat}:
	\begin{enumerate}
		\item For fixed $\{\mathbf{w}_k\}$, compute the MMSE receivers $\{\mathbf{u}_{k}\}$ and the corresponding MSE $\{e_k\}$ according to (\ref{eq:MSEreceiver_DataSharing_MIMO}) and (\ref{eq:MSE_DataSharing_MIMO});
		\item Update the MSE weights $\{\rho_k\}$ according to (\ref{eq:rho_DataSharing});
		\item For fixed $\{u_{k}\}$, $\{\rho_k\}$, and $\{ \hat{R}_k \}$ in~(\ref{QCQP_bckhaulConstraint_MIMO}), find the optimal transmit beamformers $\{w_{l,k}\}$ by solving (\ref{QCQP_DataSharing_MIMO});
		\item Update $\{\beta_{l,k}\}$ as in~(\ref{UpdateBeta_MIMO});
		\item Compute the achievable rates $\{ R_k \}$ according to (\ref{eq:Rk_DataSharing_MIMO}). Update $\hat{R}_k = R_k$, $k \in \mathcal{K}$.
	\end{enumerate}
	{\bf Until} convergence
	\caption{Weighted sum rate maximization for the data-sharing strategy}
	\label{alg:LCSparse_MIMO}
\end{algorithm}
The quantities used in the WMMSE approach in the Algorithm~\ref{alg:LCSparse_MIMO} are as follows. Let
\begin{equation}
	\mathbf{V}_k =  \Gamma_m \left ( \sum_{j \neq k} \mathbf{H}_{k} \mathbf{w}_j \mathbf{w}_{j}^{H} \mathbf{H}_{k}^{H} + \sigma^2 \mathbf{I} \right)  +  \mathbf{H}_{k} \mathbf{w}_k \mathbf{w}_{k}^{H} \mathbf{H}_{k}^{H}.
\end{equation}
\begin{itemize}
	\item The mean square error (MSE) for user $k$ is defined as
		\begin{align}
		e_k  = \mathbf{u}_{k}^{H} \mathbf{V}_k \mathbf{u}_{k} - 2 \text{Re} \left\{ \mathbf{u}_{k}^{H} \mathbf{H}_{k} \mathbf{w}_k\right\} + 1. \label{eq:MSE_DataSharing_MIMO}
		\end{align}
	\item The optimal MSE weight $\rho_k$ under fixed $\{\mathbf{w}_k\}$ and $\{\mathbf{u}_k\}$ is given by
	\begin{equation} \label{eq:rho_DataSharing}
	\rho_k = e_k^{-1}.
	\end{equation}
	\item The optimal receive beamformer $\mathbf{u}_k$ under fixed $\{\mathbf{w}_k\}$ and $\{\rho_k\}$ is given by
	\begin{equation}
	\mathbf{u}_k = \mathbf{V}_k^{-1} \mathbf{H}_{k} \mathbf{w}_k.
	\label{eq:MSEreceiver_DataSharing_MIMO}
	\end{equation}
	\item The optimal transmit beamformers $\{ \mathbf{w}_k \}$ under fixed $\{ \mathbf{u}_k \}$, $\{ \rho_k \}$ and fixed $\{ \hat{R}_k \}$ can be obtained by solving the following QCQP problem:
	\begin{subequations} \label{QCQP_DataSharing_MIMO}
		\begin{align}
		\mini_{\{\mathbf{w}_{l,k}\}} \quad & \sum_{k=1}^K \mathbf{w}_k^H \mathbf{A}_k \mathbf{w}_k-~\text{Re}\{\mathbf{b}_k^H \mathbf{w}_k\} \label{eq:obj_QCQP_DataSharing_MIMO} \\
		\sbto \quad &  \sum_{k=1}^K \left \vert w_{l,k}^m\right \vert^{2} \leq P_l^m, \quad l \in \mathcal{L}, m \in \mathcal{M} \\
		& \sum_{k=1}^K  \beta_{l,k}\hat{R}_k \left\|\mathbf{w}_{l,k}\right\|_2^2 \leq C_l, \quad l \in \mathcal{L}, \label{QCQP_bckhaulConstraint_MIMO}
		\end{align}
	\end{subequations}
	where $ \{ \mathbf{A}_k \} \in \mathbb{C}^{LM \times LM} $ and $\{\mathbf{b}_k \} \in \mathbb{C}^{LM \times 1}$ are defined to be
	\begin{align}\label{A_k_MIMO}
	\mathbf{A}_k &= \sum_{j \neq k}  \alpha_j \rho_j \Gamma_m \mathbf{H}_{j}^H \mathbf{u}_j \mathbf{u}_j^H \mathbf{H}_{j} + \alpha_k \rho_k \mathbf{H}_{k}^H \mathbf{u}_{k} \mathbf{u}_{k}^H \mathbf{H}_{k},\\
	\label{b_k_MIMO}
	\mathbf{b}_k &= 2\alpha_k \rho_k \mathbf{H}_{k}^H \mathbf{u}_k.
	\end{align}
\end{itemize}

\section{Compression Strategy}
\label{sec:compression}

\begin{figure}[t]
	\centering
	\includegraphics[width=1.0\columnwidth]{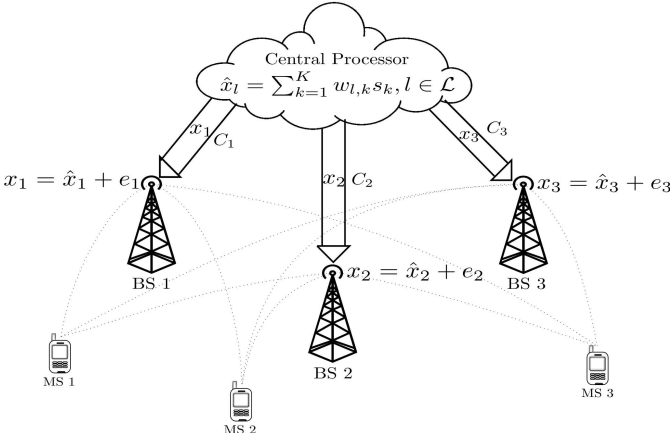}
	\caption{The compression strategy for the downlink C-RAN.}
	\label{fig:figureCompression}
\end{figure}

In the compression strategy, as shown in Fig.~\ref{fig:figureCompression}, the central processor computes the beamformed analog signals
to be transmitted by the BSs. These signals have to be compressed before they can be forwarded to the corresponding BSs through the finite-capacity backhaul links. The process of compression introduces quantization noises; the quantization noise levels depend on the backhaul capacities.

\subsection{Optimization Framework}
In the data-sharing strategy, the beamformed signal is computed at the BSs. In the compression strategy,
the beamformed signal is computed at the central processor, then compressed, sent over the backhaul links, and reproduced by the BSs.
Let $\hat{\mathbf{x}}_l \in \mathbb{C}^{M \times 1} = [\hat{x}_l^1,\ldots,\hat{x}_l^M]^T$ denote the precoded signal computed at the central processor intended for BS $l$ and $\hat{\mathbf{x}} \in \mathbb{C}^{LM \times 1} = [\hat{\mathbf{x}}_1^T,\ldots,\hat{\mathbf{x}}_L^T]^T$ be the aggregate signal intended for all the BSs. As in the data-sharing strategy, let the beamforming vector from BS $l$ to user $k$ be denoted by $\mathbf{w}_{l,k} \in \mathbb{C}^{M \times 1} = [w_{l,k}^1,\ldots,w_{l,k}^M]^T$ with $w_{l,k}^m$ being the beamforming coefficient from the $m$th antenna of BS $l$ to user $k$ and $\mathbf{w}_k \in \mathbb{C}^{LM \times 1} = [\mathbf{w}_{1,k}^T,\ldots,\mathbf{w}_{L,k}^T]^T$ be the aggregate network-wide beamformer to user $k$ from all the BSs. We can then write $\hat{\mathbf{x}}$ as
\begin{equation}
\label{eq:beamformedSignal_Compression}
\hat{\mathbf{x}}=\sum_{k=1}^K \mathbf{w}_k s_k.
\end{equation}
The analog signal $\hat{\mathbf{x}}$ is then compressed and forwarded to BSs. We model the quantization process as
\begin{equation}
\label{eq:quantizationModel}
\mathbf{x} = \hat{\mathbf{x}}  + \mathbf{e},
\end{equation}
where the quantization noise $\mathbf{e}$ is assumed to be complex Gaussian distributed with zero mean and covariance matrix $\mathbf{Q} \in \mathbf{C}^{LM \times LM}$ and is independent of $\hat{\mathbf{x}}$. Under this model, the achievable rate for user $k$, accounting for the SNR gap, is given by
\begin{equation}
\label{eq:Rk_Compression_MIMO}
R_k = \log \left( 1 +\frac{\text{SINR}_k}{\Gamma_m} \right),
\end{equation}
where the SINR at user $k$ is
\begin{equation}
	\text{SINR}_k = \mathbf{w}_{k}^{H} \mathbf{H}_{k}^{H} \left (\sum_{j \neq k } \mathbf{H}_{k} \mathbf{w}_j \mathbf{w}_{j}^{H} \mathbf{H}_{k}^{H} + \sigma^2 \mathbf{I} + \mathbf{H}_k \mathbf{Q}\mathbf{H}_k^H \right )^{-1} \mathbf{H}_{k} \mathbf{w}_k.
\end{equation}
We consider independent quantization at each antenna at all the BSs,
in which case $\mathbf{Q}$ is a diagonal matrix with diagonal entries $q_l^m$.
(Multivariate compression is also possible and has been studied in
\cite{ParkSimeoneSahinShamai13}.)
Assuming an ideal vector quantizer, the
quantization noise level $q_l^m$ and the backhaul capacity $C_l^m$ allocated to each antenna at each BS is related as (from rate-distortion theory~\cite{berger1971rate})
\begin{equation}
\label{eq:rate_distortion}
\log \Bigg( 1 +  \frac{\sum_{k=1}^K{\left \vert w_{l,k}^m \right \vert^2}}{q_l^m} \Bigg) \le C_l^m.
\end{equation}
However, the quantizers used in practice for compression can be far from ideal. In order to capture these losses, we introduce a notion of gap to rate-distortion limit. Following~\cite{GrayNeuhoff_TIT98}, we note that the operational distortion, $\delta(R)$, achieved by virtually all practical quantizers at high resolution follows the relation
\begin{equation}
\delta(R) = \Gamma_q {\rm var}(X) 2^{-R},
\end{equation}
where ${\rm var}(X)$ is the variance of the signal being quantized, $R$ is the rate of the quantizer, and $\Gamma_q$ is a constant that depends on the particular choice of the quantizer.
For example, for a fixed-rate (uncoded) uniform scalar quantizer, $\Gamma_q$ = $\frac{\sqrt{3}\pi}{2}$, which is approximately 2.72.
For a uniform scalar quantizer followed by variable-rate entropy coding, $\Gamma_q = \frac{\pi e}{6}$, which is approximately 1.42.
Note that $\Gamma_q = 1$ corresponds to the distortion achievable by the best possible vector quantization scheme. Accounting for this, we can rewrite the relation (\ref{eq:rate_distortion}) above as
\begin{equation}\label{quantization_relation}
\log \Bigg( 1 +  \frac{\Gamma_q\sum_{k=1}^K{\left \vert w_{l,k}^m \right \vert^2}}{q_l^m} \Bigg) \le C_l^m.
\end{equation}

The quantization noise relation described by (\ref{quantization_relation}) assumes that individual BSs have access to the quantization codebooks used at the central processor for compressing the signals intended for all of their antennas. The quantization codebooks depend on the variance of the signal being compressed $\sum_{k=1}^K{\vert w_{l,k}^m \vert^2}$ and rate of the quantizer $C_l^m$. Since we are also designing the beamforming coefficients $\{w_{l,k}\}$ at each user scheduling time slot, the variance of the signal being compressed can change at each user scheduling iteration. Also, the rate of the quantizer, $C_l^m$, used for compressing the signal of antenna $m$ at BS $l$, depends on the backhaul capacity allocated to antenna $m$ of BS $l$. This allocation can also potentially change from time slot to time slot. Thus, to achieve (\ref{quantization_relation}), the information about the quantization codebooks used at the central processor for all antennas of a BS needs to be sent to that BS at the start of each user scheduling iteration.

In practice, however, it may not be feasible to convey all such relevant codebook information from the central processor to each individual BS at each user scheduling time slot. We consider below two optimization formulations, one that allows for adaptive quantization codebooks, while the other with fixed quantization codebooks, and develop the corresponding algorithms in each case.

\subsection{Optimization Methodology}

\subsubsection{Adaptive Quantization}
We refer to the situation when the quantization codebooks are allowed to be
changed at the central processor at each user scheduling time slot as adaptive
quantization. It is adaptive in the sense that, depending on the active users
and their priority weights, the quantization codebooks are allowed to be
adapted. Recall that the quantization codebooks depend on the variance of the
signal being compressed $\sum_{k=1}^K{\vert w_{l,k}^m \vert^2}$ and the rate of
the quantizer $C_l^m$. The variance of the signal to be compressed depends on
the beamformers which can change per each time slot; the rate of the
quantizer depends on the backhaul capacity allocated for compression. In the
case of single-antenna BSs, since the backhaul capacity per-antenna is
fixed (which is same as the per-BS backhaul capacity), the rate of the
quantizer for that BS is fixed. In the case of multiple antennas, the rate of
the quantizers used for different antennas at a BS can potentially be
different, but for simplicity, this paper assumes that the compression rate
allocation among different antennas at a BS to be uniform among all the antennas, 
i.e., 
\begin{equation}
C_l^m = \frac{C_l}{M}, \quad l \in \mathcal{L}, m \in \mathcal{M} .
\end{equation}

The design of the compression strategy with adaptive compression can now be stated
as a weighted sum rate maximization problem over the transmit
beamformers and the quantization noise levels as follows:
\begin{subequations} \label{eq:optimizationCompression_MIMO_adaptive}
	\begin{align}
	\underset{\{\mathbf{w}_{l,k}\}, \{q_l^m\}}{\textrm{maximize}}  \quad & \sum_{k=1}^K \alpha_k R_k  \label{obj_compression_MIMO_adative]} \\
	\textrm{subject to} \quad & \sum_{k=1}^K{\left \vert w_{l,k}^m \right \vert^2} +q_l^m \le P_l^m, \quad l \in \mathcal{L}, m \in \mathcal{M} \label{quantizationLowerBound_MIMO_adaptive} \\
	\quad & \sum_{k=1}^K{\left \vert w_{l,k}^m \right \vert^2} - \frac{2^{C_l^m}-1}{\Gamma_q}~q_l^m \le 0, ~ l \in \mathcal{L}, m \in \mathcal{M}, \label{quantizationUpperBound_MIMO_adaptive}
	\end{align}
\end{subequations}
where the constraint (\ref{quantizationUpperBound_MIMO_adaptive}) is just a reformulation of (\ref{quantization_relation}), while the constraint (\ref{quantizationLowerBound_MIMO_adaptive}) is the per-antenna power constraint at BS $l$.

Finding the globally optimal solution to (\ref{eq:optimizationCompression_MIMO_adaptive}) is challenging. An iterative approach based on the majorize-minimization (MM) algorithm has been suggested in \cite{ParkSimeoneSahinShamai13}. The algorithm in \cite{ParkSimeoneSahinShamai13} transforms $\mathbf{w}_k\mathbf{w}_k^H$ into a non-negative definite matrix variable $\mathbf{R}_k$ and ignores the rank constraint on $\mathbf{R}_k$ in the optimization.
In this paper, we propose a novel way to solve (\ref{eq:optimizationCompression_MIMO_adaptive}) by reformulating it as an equivalent WMMSE problem then using the block coordinate descent method between the groups of variables of the transmit beamformers $\{ \mathbf{w}_k \}$ and the quantization noise levels $\{ q_l^m \}$, the receive beamformers $\{ u_k \}$, and the MSE weights $\{ \rho_k \}$. We summarize the overall algorithm to solve (\ref{eq:optimizationCompression_MIMO_adaptive}) in Algorithm \ref{alg:pureCompression}. The algorithm can be shown to reach a stationary point of (\ref{eq:optimizationCompression_MIMO_adaptive}).
Following a similar analysis as in the data-sharing strategy, the computational complexity of the algorithm scales as $O((KLM)^{3.5}T)$, where $T$ is the number of iterations to converge.
\begin{algorithm}[t]
	{\bf Initialization}: $\{\mathbf{w}_{k}\}, \{ q_l^m \}$; \\
	{\bf Repeat}:
	\begin{enumerate}
		\item For fixed $\{\mathbf{w}_{k}\}, \{ q_l \}$, compute the MMSE receivers $\{\mathbf{u}_{k}\}$ and the corresponding MSE $\{e_k\}$ according to (\ref{eq:MSEreceiver_Compression_MIMO}) and (\ref{eq:MSE_Compression_MIMO});
		\item Update the MSE weights $\{\rho_k\}$ according to (\ref{eq:rho_DataSharing});
		\item For fixed $\{\mathbf{u}_{k}\}$ and $\{\rho_k\}$, find the optimal transmit beamformers $\{\mathbf{w}_{k}\}$ and quantization noise levels $\{q_l^m\}$ by solving the convex optimization problem~(\ref{eq:QCQP_Compression});
	\end{enumerate}
	{\bf Until} convergence
	\caption{Weighted sum rate maximization for the compression strategy with adapative quantization}
	\label{alg:pureCompression}
\end{algorithm}

The quantities used in the WMMSE approach in the Algorithm \ref{alg:pureCompression} are as follows. Let
\begin{equation}
	\mathbf{V}_k = \Gamma_m \left ( \sum_{j \neq k} \mathbf{H}_{k} \mathbf{w}_j \mathbf{w}_{j}^{H} \mathbf{H}_{k}^{H} + \sigma^2 \mathbf{I} +\mathbf{H}_k \mathbf{Q}\mathbf{H}_k^H \right)  +  \mathbf{H}_{k} \mathbf{w}_k \mathbf{w}_{k}^{H} \mathbf{H}_{k}^{H}.
\end{equation}
\begin{itemize}
	\item The MSE for user $k$ is calculated as
	\begin{equation}
	\label{eq:MSE_Compression_MIMO}
	e_k  = \mathbf{u}_{k}^{H} \mathbf{V}_k \mathbf{u}_{k} - 2 \text{Re} \left\{ \mathbf{u}_{k}^{H} \mathbf{H}_{k} \mathbf{w}_k\right\} + 1.
	\end{equation} 
	\item The optimal receive beamformer $\mathbf{u}_k$ under fixed $\{\mathbf{w}_k\}$ and $\{\rho_k\}$ is given by
	\begin{equation}
\mathbf{u}_k = \mathbf{V}_k^{-1} \mathbf{H}_{k} \mathbf{w}_k.
\label{eq:MSEreceiver_Compression_MIMO}
\end{equation}
	\item The optimization of the transmit beamformers $\{ \mathbf{w}_{k} \}$ and the quantization noise levels $\{ q_l^m \}$ under fixed $\{\mathbf{u}_{k}\}$ and $\{\rho_k\}$ is solved via the following convex program:
	\begin{subequations} \label{eq:QCQP_Compression}
		\begin{align}
		\underset{\{\mathbf{w}_{l,k}\},\{q_l^m\}}{\textrm{minimize}} \label{obj_QCQP_Compression}
		\quad & \sum_{k=1}^K \mathbf{w}_k^H \mathbf{A}_k \mathbf{w}_k -~\text{Re} \{ \mathbf{b}_{k}^H \mathbf{w}_k\} \nonumber \\ & \qquad + \Gamma_m \alpha_k \rho_k  \mathbf{u}_k^H\mathbf{H}_k\mathbf{Q}\mathbf{H}_k^H\mathbf{u}_k \\
		\textrm{subject to}
		\quad & \sum_{k=1}^K \left \vert w_{l,k}^m \right \vert^{2} + q_l^m \leq P_l^m, \quad l \in \mathcal{L}, m \in \mathcal{M} \label{QCQP_quantizationUpperBound}\\
		\quad & \sum_{k=1}^K{\left \vert w_{l,k}^m \right \vert^2} - \frac{2^{C_l^m}-1}{\Gamma_q}~q_l^m \le 0,~ l \in \mathcal{L}, m \in \mathcal{M}, \label{QCQP_quantizationLowerBound}
		\end{align}
	\end{subequations}
	where $\{\mathbf{A}_k\}$ and $\{\mathbf{b}_k\}$ are as defined in (\ref{A_k_MIMO}) and (\ref{b_k_MIMO}), respectively.
	We further observe that the convex optimization problem~(\ref{eq:QCQP_Compression}) has a particular structure that can be exploited. Observe that the two constraints~(\ref{QCQP_quantizationUpperBound}) and~(\ref{QCQP_quantizationLowerBound}) provide an upper and a lower bound on $\{ q_l^m \}$, respectively. Since the objective~(\ref{obj_QCQP_Compression}) is monotonically decreasing in $\{ q_l^m \}$, we can replace the inequality with equality in the constraint~(\ref{QCQP_quantizationLowerBound}) and substitute $\{ q_l^m \}$ from~(\ref{QCQP_quantizationLowerBound}) into the objective~(\ref{obj_QCQP_Compression}) and the constraint~(\ref{QCQP_quantizationUpperBound}). This results in a QCQP problem in only a single set of variables $\{ \mathbf{w}_{l,k} \}$, which can be solved efficiently by standard solvers such as CVX \cite{cvx}.
\end{itemize}

\subsubsection{Fixed Quantization}
We now consider the quantization model when the quantization codebooks are
fixed at the central processor and at the BSs. The achievable rate is as given
by (\ref{eq:Rk_Compression_MIMO}). To fix the codebook for the quantizer for
antenna $m$ of BS $l$, we assume that the range of the quantizer is constrained
within the power constraint for the antenna, $P_l^m$. Further, we assume that the
allocation of the backhaul capacity to each antenna of a BS is uniform, i.e., 
$C_l^m = \frac{C_l}{M},~ l \in \mathcal{L}, m \in \mathcal{M}.$
With these assumptions, the quantization relation (\ref{quantization_relation}) becomes
\begin{equation}\label{quantization_relation_MIMO}
\log \Bigg( 1 +  \frac{\Gamma_q P_l^m}{q_l^m} \Bigg) \le C_l^m.
\end{equation}
We can now formulate the weighted sum rate maximization problem over the transmit
beamformers and the quantization noise levels as:
\begin{subequations} \label{eq:optimizationCompression_MIMO}
	\begin{align}
	\underset{\{\mathbf{w}_{l,k}\}, \{q_l^m\}}{\textrm{maximize}}  \quad & \sum_{k=1}^K \alpha_k R_k  \label{obj_compression_MIMO} \\
	\textrm{subject to} \quad & \sum_{k=1}^K{\left \vert w_{l,k}^m \right \vert^2} +q_l^m \le P_l^m, \quad l \in \mathcal{L}, m \in \mathcal{M} \label{quantizationLowerBound_MIMO} \\
	\quad & q_l^m \ge \frac{\Gamma_q P_l^m} {2^{C^m_l} -1} , \quad l \in \mathcal{L}, m \in \mathcal{M}, \label{quantizationUpperBound_MIMO}
	\end{align}
\end{subequations}
where the constraint (\ref{quantizationUpperBound_MIMO}) is a reformulation of (\ref{quantization_relation_MIMO}). 

In order to solve the optimization problem (\ref{eq:optimizationCompression_MIMO}), we first observe that the objective (\ref{obj_compression_MIMO}) is a decreasing function of $q_l^m$. The constraint (\ref{quantizationUpperBound_MIMO}) provides a lower bound on $q_l^m$, while the constraint (\ref{quantizationLowerBound_MIMO}) provides an upper bound. Hence, the constraint (\ref{quantizationUpperBound_MIMO}) will always be met with equality at a stationary point. Thus, we can substitute the value of $q_l^m$ from (\ref{quantizationUpperBound_MIMO}) into the objective (\ref{obj_compression_MIMO}) as well as the constraint (\ref{quantizationLowerBound_MIMO}) and eliminate the variables $q_l^m$. This modifies the constraint (\ref{quantizationLowerBound_MIMO}) into
\begin{equation}
\label{ModifiedConstraint_Compression_MIMO}
\sum_{k=1}^K{\left \vert w_{l,k}^m \right \vert^2} \le P_l^m \left(1 - \frac{\Gamma_q}{2^{C_l^m} - 1}\right), \quad l \in \mathcal{L}, m \in \mathcal{M}.
\end{equation}
We then end up with a weighted sum rate maximization problem over only the
beamformers $\{\mathbf{w}_{l,k}\}$ with modified per-antenna power constraints,
which can be tackled by solving its equivalent WMMSE problem.

\section{Performance Comparison of Data-sharing versus Compression Strategies}
\label{sec:data_sharing_vs_compression}

A main contribution of the first part of this paper is a comparison
between data-sharing and compression strategies under the same network
utility maximization framework using a unified WMMSE approach. Toward
this end, we consider a 7-cell wrapped-around two-tier heterogeneous network with intercell distance of 0.8 km over a 10 MHz channel bandwidth.
Each cell is a regular hexagon with 1 macro-BS at the center and 3 pico-BSs equally separated in space. There are 30 users randomly placed in each cell.
The power budget per each antenna is 43 dBm at macro-BSs and 30 dBm at pic-BSs. We assume antenna gain of 15 dBi.
The path loss in dB from macro-BS to user is modeled as $128.1+ 37.6 \log_{10}(d)$, while from pico-BS to user as $140.7+ 36.7 \log_{10}(d)$, with log-normal shadowing of 8 dB and small scale Rayleigh fading of 0 dB. The combined background noise and  interference caused by two tiers of cells outside the 7-cells is estimated to be at -150 dBm/Hz. We assume an SNR gap of $\Gamma_m=9$ dB (corresponding to uncoded QAM transmission) and a gap to rate-distortion limit of $\Gamma_q=4.3$ dB (corresponding to uncoded fixed-rate uniform scalar quantizer).
At each time slot, we solve the respective network optimization problems and
update the weights in the weighted sum rate maximization according to the proportional fair criterion.

\begin{figure}[t]
	\centering
	\includegraphics[width=0.96\columnwidth]{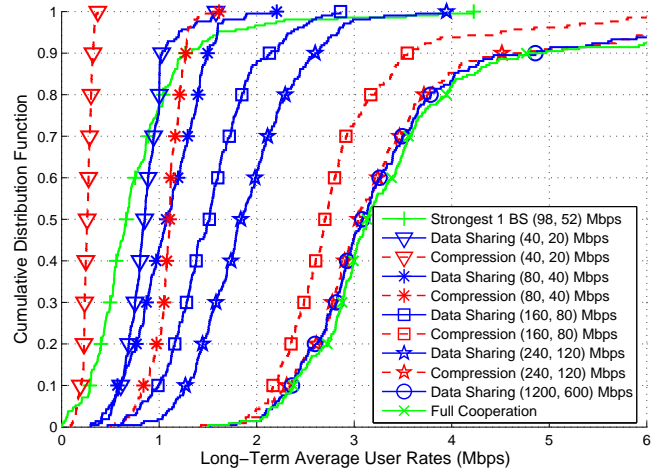}
	\caption{Cumulative distribution function of user rates for the
data-sharing and compression strategies with adaptive quantization and single antenna terminals for varying
fronthaul/backhaul capacities.}
	\label{fig:cdfComparison_new}
\end{figure}

In the first set of simulations, we compare the performance of the data-sharing
strategy and the compression strategy with adaptive quantization and with
single transmit antenna at both the macro-BSs and pico-BSs, and single receive
antenna at the users. Fig.~\ref{fig:cdfComparison_new} shows the cumulative
distribution of user rates under varying fronthaul/backhaul capacities for both
strategies.  Plots for the compression strategy are shown in red, while those
for data-sharing strategy are shown in blue.  For reference, we also include
the full cooperation case with infinite backhaul capacity and the baseline
scheme of no cooperation with each user connected to the strongest BS.

When the fronthaul/backhaul capacity is low at 40 Mbps/macro-BS and 20
Mbps/pico-BS, the data-sharing strategy outperforms the compression strategy.
The 50-percentile rate for the data-sharing strategy is about 3 times that of
the compression strategy. If we double the fronthaul/backhaul capacity to 80
Mbps/macro-BS and 40 Mbps/pico-BS, the compression strategy becomes comparable
to the data-sharing strategy and both have about the same 50-percentile user
rates. At this operating point, the sum fronthaul/backhaul capacity is about 6
times that of the average sum rate per cell. We also observe that the
compression strategy favors low rate users while the data-sharing strategy
favors high rate users. A reason for this is that the compression strategy
under low fronthaul capacity is limited by the quantization noises which are
about the same for all the BS signals resulting in more uniform user rates.

We observe that with moderate-to-high fronthaul/backhaul capacity of 160 Mbps/macro-BS
and 80 Mbps/pico-BS, the compression strategy outperforms the data-sharing
strategy with the 50-percentile rate for the compression strategy more than 2.5
times than that of data-sharing. Increasing the fronthaul/backhaul in this regime
improves the compression strategy drastically, while the data-sharing strategy
sees only a moderate increase. This is because at low backhaul capacity, the
performance of the compression strategy is limited by the quantization noises.
An increase in fronthaul capacity reduces the quantization noise levels
exponentially, while a similar increase in the backhaul capacity does not buy
as much for the data-sharing strategy. Finally with a fronthaul of 240
Mbps/macro-BS and 120 Mbps/pico-BS, the compression strategy performs close to
the full cooperation limit, while for the data-sharing strategy, backhaul
capacities of 1200 Mbps/macro-BS and 600 Mbps/pico-BS are needed to get as
close. This is because to match the full cooperation limit, the data-sharing
strategy needs large cluster size, leading to significantly higher backhaul
capacity.

\begin{figure}[t]
	\centering
	\includegraphics[width=1.0\columnwidth]{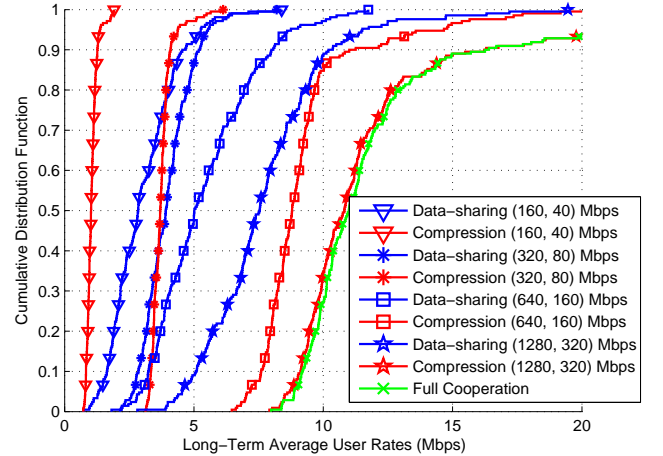}
	\caption{Cumulative distribution functions of user rates for the
data-sharing and compression strategies with fixed quantization and multi-antenna terminals for varying fronthaul/backhaul capacities.}
	\label{fig:cdfComparison_DataSharing_Compression_MIMO_FullCSI}
\end{figure}

In the second set of simulations, we compare the performance of the
data-sharing strategy and the compression strategy with fixed quantization when
all the terminals have multiple antennas. We assume 4 antennas per macro-BS, 2
antennas per pico-BS, and 2 receive antennas for each user.
Fig.~\ref{fig:cdfComparison_DataSharing_Compression_MIMO_FullCSI} shows the
cumulative distribution of user rates with varying fronthaul/backhaul
capacities for both strategies. For reference, the plot for full cooperation
with infinite fronthaul/backhaul capacity is also included. We observe similar
trends as in the case of single-antenna terminals and adaptive quantization.
When the backhaul capacity is low at 160 Mbps/macro-BS and 40 Mbps/pico-BS
(note that on average per-antenna fronthaul/backhaul capacities are maintained
at 40 Mbps/macro-BS antenna and 20 Mbps/pico-BS antenna), the data-sharing
strategy outperforms the compression compression strategy. The 50-percentile
rate for the data-sharing strategy is about 2.5 times that of the compression strategy.

If we double the fronthal/backhaul capacity to 320 Mbps/macro-BS and 80
Mbps/pico-BS, we see that the compression strategy becomes comparable to the
data-sharing strategy and both have about the same 40-percentile user rates. 
In this regime the sum fronthaul/backhaul capacity is about 5 times that of the
average sum rate per cell. This is in the similar range to what we observe in
the single-antenna case. As the fronthaul/backhaul capacity is increased
further to 640 Mbps/macro-BS and 160 Mbps/pico-BS, the compression strategy
starts to significantly outperform the data-sharing strategy with the
50-percentile user rate being about 80\% higher than that in the data-sharing
strategy. With the fronthaul/backhaul capacity of 1280 Mbps/macro-BS and 320
Mbps/pico-BS, the compression strategy already achieves the maximum achievable
rates of the full cooperation. At this fronthaul/backhaul capacity the
quantization noises are small enough that they do not affect the user rates. At
the same backhaul capacity, the data-sharing strategy is still far behind that
of full cooperation. This is because the backhaul capacity is not high enough
to allow for the backhaul exchange required to maintain full cooperation.

\begin{figure}[t]
	\centering
	\includegraphics[width=1.0\columnwidth]{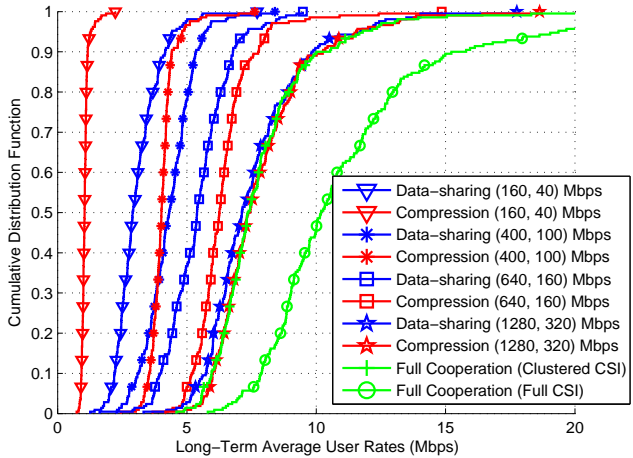}
	\caption{Comparison of cumulative distribution functions of user rates
for the data-sharing and compression strategies with clustered CSI for varying fronthaul/backhaul capacities.}
	\label{fig:cdfComparison_DataSharing_Compression_MIMO_LimitedCSI}
\end{figure}

It is important to note that the benefits from the compression strategy come at a cost of high CSI requirements at the central processor. To understand the impact of CSI on the data-sharing and compression strategies, we limit the amount of CSI available at the central processor by only allowing CSI of the few strongest BSs for each user. We call such a restriction clustered CSI, because the CSI of only a cluster of BSs around any user is available. All the proposed algorithms can be adapted when such clustered CSI is available for each user.

Fig.~\ref{fig:cdfComparison_DataSharing_Compression_MIMO_LimitedCSI} shows the
cumulative distribution of user rates for both strategies when the CSI is
limited to only 7 strongest BSs for each user. We observe that the general
trend seen in the above two cases remains the same. At low fronthaul/backhaul
capacity of 160 Mbps/maco-BS and 40 Mbps/pico-BS, the data-sharing strategy
outperforms the compression strategy, while at the high fronthaul/backhaul
capacity of
640 Mbps/macro-BS and 160 Mbps/pico-BS, the compression strategy outperforms
the data-sharing strategy. However, notice that, in this case the compression
strategy is not as significantly better than the data-sharing strategy as in the previous case. The
50-percentile user rate of the compression strategy is only 20\% better than
that of the data-sharing strategy, as compared with the case with full CSI when 
it is almost 80\% better.
Finally, when the fronthaul/backhaul capacity is high at 1280 Mbps/macro-BS and 320
Mbps/pico-BS, both the data-sharing and compression strategies saturate to the
full cooperation user rates with infinite backhaul capacity under clustered CSI.
For reference, the plot with full cooperation with infinite fronthaul/backhaul capacity
and full CSI is also included to highlight the performance loss that is
attributed to the lack of full CSI.

\begin{figure}[t]
	\centering
	\includegraphics[width=1.0\columnwidth]{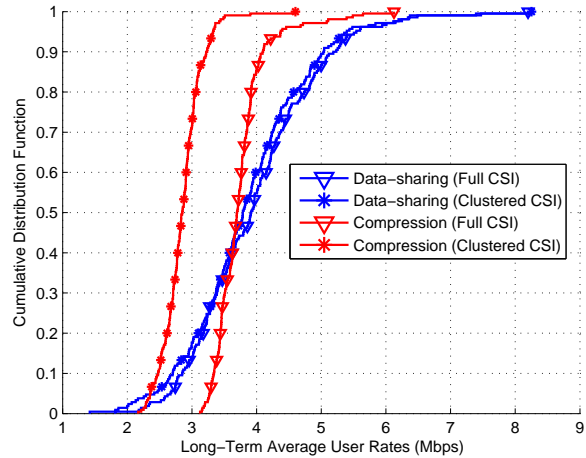}
	\caption{Comparison of the cumulative distribution functions of user
rates for the data-sharing and compression strategies with full CSI and clustered CSI at the fronthaul/backahul capacity of 320 Mbps/macro-BS and 80
Mbps/pico-BS.}
	\label{fig:cdfComparison_DataSharing_Compression_MIMO_LimitedCSIvsFullCSI}
\end{figure}

In order to closely look at the how the lack of full CSI affects the
data-sharing and compression strategies, we fix the fronthaul/backhaul capacity
at 320 Mbps/macro-BS and 80 Mbps/pico-BS. This is the regime where the two
strategies are comparable in the case with full CSI.
Fig.~\ref{fig:cdfComparison_DataSharing_Compression_MIMO_LimitedCSIvsFullCSI}
shows the cumulative distribution of user rates for both strategies with full
CSI and with clustered CSI at this fixed fronthaul/backhaul capacity. As is
evident from the plot, the compression strategy suffers more than the
data-sharing strategy when only clustered CSI is available. The reason for this
behavior is that the compression strategy benefits from having the ability to
fully cooperate at the central processor, but when only clustered CSI is
available at the central processor, the cooperation cluster size at the central
processor becomes limited. The data-sharing strategy on the other hand does not
pay as much penalty because the cooperation cluster size for the data-sharing
strategy is already small due to the backhaul capacity limitations. As a
result, we also see that the fronthaul/backhaul capacity at which the two
strategies are comparable is higher when only clustered CSI is available (at
400 Mbps/macro-BS and 100 Mbps/pico-BS) than the case with full CSI (at 320
Mbps/macro-BS and 80 Mbps/pico-BS).

To conclude, the data-sharing and compression strategies show different
characteristics depending on the operating regime of interest. The data-sharing
strategy is advantageous when the fronthaul/backhaul is severely limited as the
compression strategy suffers from high quantization noises introduced under low
fronthaul capacity, while the compression strategy is advantageous when the
fronthaul/backhaul capacity is large due to its ability to have full
cooperation before quantization. Further, the compression strategy is more
sensitive to the availability of CSI than the data-sharing strategy, as in the
former the benefits stem from the ability to fully cooperate at the central
processor, which is affected adversely by the lack of full CSI.

\section{Hybrid Data-Sharing and Compression Strategy}
\label{sec:hybrid}

In the data-sharing based cooperation scheme, the backhaul links
are exclusively used to carry user messages. The advantage of such an
approach is that BSs get clean messages which they can use for joint
encoding. However, the backhaul capacity constraint limits the
cooperation cluster size for each user. In the compression based scheme,
the precoding operation is exclusively performed at the central
processor. The main advantage of such an approach is that, since the
central processor has access to all the user data, it can form a joint
precoding vector using all the user messages, thus achieving full BS
cooperation.  Additionally, the BSs can now be completely oblivious of
the user codebooks as the burden of preprocessing is shifted from the
BSs to the central processor.  However, since the precoded signals are
compressed, we pay a price in the form of quantization noises.

The second part of this 
paper proposes a hybrid compression and data-sharing strategy
in which the precoding operation is split between the central
processor and the BSs. The rationale is that as the desired
precoded signal typically consists of both strong and weak users, it may be beneficial to send clean messages for the strong users, rather than including them as a part of the signal to be compressed.
In so doing, the amplitude of the signal that needs to be compressed
can be lowered, and the required number of compression bits reduced.

Building on this intuition, we propose an approach where a
part of backhaul capacity is used to send direct messages for some
users (for whom the BSs are better off receiving messages directly,
instead of their contributions in the compressed
precoded signals) and the remaining backhaul capacity is used
to carry the compressed signal that combines the contributions from the
rest of the users. Typically, each BS receives direct messages for
the strong users and compressed precoded signals combining messages
of the rest of the weak users in the network. Each BS then combines
the direct messages with the decompressed signal, and transmits the
resulting precoded signal on its antenna.
Note that the appropriate beamforming coefficients are assumed to be
available at both the cloud processor and at the BSs.

We point out that a dirty-paper coding based scheme proposed in \cite{SimeoneSomekhPoorShamai09} also
makes use of the backhaul links to carry a combination of user message and the
compressed version of interfering signal from the neighboring BS in a
simplified linear array model. But the scheme
of~\cite{SimeoneSomekhPoorShamai09} is limited to the simplified linear array model;
it also does not provide a method to decide if and what user messages should be
shared among the BSs and what signals should be compressed.

\subsection{Optimization Framework}

\begin{figure}[t]
	\centering
	\includegraphics[width=1.0\columnwidth]{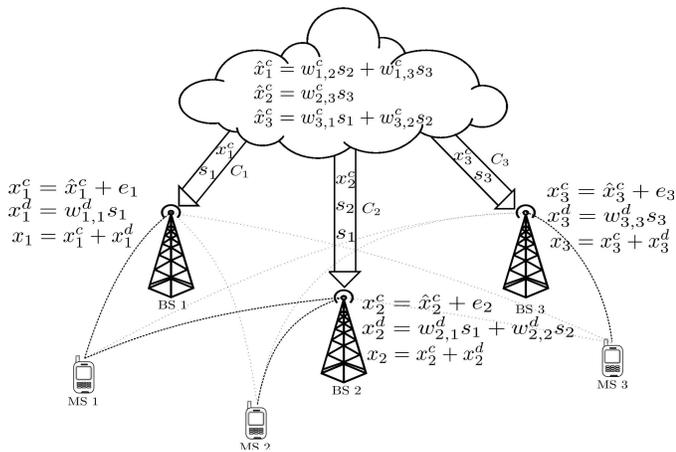}
	\caption{The hybrid data-sharing and compression strategy for the downlink C-RAN.}
	\label{fig:figureHybrid_explicit}
\end{figure}

A key question is how to optimize the hybrid data-sharing and
compression strategy.
In the hybrid strategy, as shown in Fig.~\ref{fig:figureHybrid_explicit}, the central processor computes a part of the beamformed analog signals to be transmitted by BSs. These signals are compressed and sent over to BSs using a part of the backhaul capacity. For the rest of the beamformed signal, the central processor sends digital data of selected users to the BSs using the remaining backhaul capacity. To simplify the description of the hybrid strategy, we assume single-antenna at the BSs and the user terminals.

The idea is to introduce separate beamforming coefficients for the data-sharing and compression parts. Let $\mathbf{w}_k^c \in \mathbb{C}^{L \times 1} = [{w}^c_{1,k},\ldots,{w}^c_{L,k}]^T$ be the beamformers for user $k$ used to compute the beamformed signal that is going to be compressed at the central processor. Let $\hat{\mathbf{x}}^c \in \mathbb{C}^{L \times 1} = [\hat{x}_1^c,\ldots,\hat{x}_L^c]^T$ denote the beamformed signals intended for all the BSs computed at the central processor. These are given by
\begin{equation}
\hat{\mathbf{x}}^c=\sum_{k=1}^K \mathbf{w}_k^c s_k.
\label{eq:beamformedSignal_Compression_Hybrid}
\end{equation}
The quantization process for $\hat{\mathbf{x}^c}$ is again modeled as
\begin{equation}
\label{eq:quantization_hybrid}
\mathbf{x}^c = \hat{\mathbf{x}}^c + \mathbf{e},
\end{equation}
where $\mathbf{e}$ is the quantization noise with covariance $\mathbf{Q}
\in \mathbb{C}^{L \times L}$ assumed to be Gaussian and independent of $\hat{\mathbf{x}}^c.$
Assuming independent quantization at each BS, in which case $\mathbf{Q}$ is a diagonal matrix with diagonal entries $q_l$, the amount of backhaul capacity consumed by BS $l$, $C_l^c$, for the compression part of its total beamformed signal is given by
\begin{equation}\label{quantization_relation_Hybrid}
\log \Bigg( 1 +  \frac{\Gamma_q\sum_{k=1}^K{\left \vert w_{l,k}^c \right \vert^2}}{q_l} \Bigg) \le C_l^c.
\end{equation}
Similarly, let $\mathbf{w}_k^d \in \mathbb{C}^{L \times 1} = [{w}^d_{1,k},\ldots,{w}^d_{L,k}]^T$ be the beamformers that are used for data-sharing at the BSs and $\mathbf{x}^d \in \mathbb{C}^{L \times 1} = [x_1^d,\ldots,x_L^d]^T$ denote the beamformed signals computed at the BSs using the direct data given by
\begin{equation}
\mathbf{x^d}=\sum_{k=1}^K \mathbf{w}_k^d s_k.
\label{eq:beamformedSignal_DataSharing_Hybrid}
\end{equation}
If BS $l$ does not receive direct data for user $k$, then $w_{l,k}^d = 0$. The amount of backhaul capacity, $C_l^d$, consumed by BS $l$ for the data-sharing part is then given by
\begin{equation}
\mathbbm{1}\left\{ \left \vert w_{l,k}^d \right \vert^{2} \right\} R_k \le C_l^d,
\end{equation}
where the indicator function is used to indicate whether BS $l$ participates in
computing the beamformed signal using the direct data for user $k$. If so, the
backhaul needs to support the user rate $R_k$. Note that we neglect the portion
of the backhaul capacity that would be needed to communicate the beamforming
coefficients at the start of each user scheduling iteration as the required rate
is negligible compared to the direct data rate.
The final beamformed signal transmitted by BSs to the users, $\mathbf{x}$, is then the sum of compressed beamformed signals, $\mathbf{x}^c$, communicated through the backhaul link and the direct beamformed signal, $\mathbf{x}^d$, computed at the BSs, i.e.,
\begin{equation}
\mathbf{x} = \mathbf{x}^c + \mathbf{x}^d.
\end{equation}
The achievable rate for user $k$ is then
\begin{equation}
R_k = \log \left( 1 +  \frac{\text{SINR}_k}{\Gamma_m} \right),
\label{eq:Rk_Hybrid_separate}
\end{equation}
where the SINR at user $k$ is
\begin{equation}
	\text{SINR}_k = \frac{\left \vert \mathbf{h}_k^{H}\left(\mathbf{w}_k^c + \mathbf{w}_k^d\right) \right \vert ^2}{ \sum_{j \neq k} \left \vert \mathbf{h}_k^{H}\left(\mathbf{w}_j^c+\mathbf{w}_j^d\right) \right \vert ^2 + \sigma^2 +  \mathbf{h}_k^{H}\mathbf{Q}\mathbf{h}_k }.
\end{equation}
If we let $\mathbf{w}_k^c + \mathbf{w}_k^d = \mathbf{w}_k$, $k \in \mathcal{K}$, the rate $R_k$ can be simplified to
\begin{equation}
R_k = \log \left( 1 +  \frac{\left \vert \mathbf{h}_k^{H}\mathbf{w}_k\right \vert^2}{ \Gamma_m \left( \sum_{j \neq k} \left \vert \mathbf{h}_k^{H}\mathbf{w}_j \right \vert^2 + \sigma^2 + \mathbf{h}_k^{H}\mathbf{Q}\mathbf{h}_k \right) } \right),
\label{eq:Rk_Hybrid}
\end{equation}
where $\mathbf{w}_k \in \mathcal{C}^{L \times 1}$ can be thought of as the final combined beamformer for user $k$.

Now the weighted sum rate maximization problem for the hybrid strategy can be formulated as follows:
\begin{subequations} \label{hybrid_original_optimization}
	\begin{align}
	\maxi_{\substack{\{w_{l,k}^d\},\{w_{l,k}^c\},\\ \{w_{l,k}\},\{q_l\}}} \quad & \sum_{k=1}^K \alpha_k R_k \label{hybrid_objective} \\
	\sbto \quad &  \sum_{k=1}^K \left \vert w_{l,k} \right \vert^{2} + q_l \leq P_l, \quad l \in \mathcal{L} \\
	& \sum_{k=1}^K \mathbbm{1}\left\{ \left \vert w_{l,k}^d \right\vert^{2} \right\} R_k \nonumber \\ & \quad + \log \left( 1 +  \frac{\Gamma_q\sum_{k=1}^K{\left \vert w_{l,k}^c \right \vert^2}}{q_l} \right) \leq C_l, \quad l \in \mathcal{L}  \label{BackhaulConstraint_Hybrid} \\
	& w_{l,k}^d + w_{l,k}^c = w_{l,k}, \quad l \in \mathcal{L}, k \in \mathcal{K}.
	\end{align}
\end{subequations}

Note that in the problem formulation
(\ref{hybrid_original_optimization}) above, it may seem at first that,
we allow a more general hybrid strategy where a user $k$ can
\textit{both} participate in direct data-sharing to a BS $l$ as well
as be part of the signal compressed by that BS, if both the
beamforming coefficients $w_{l,k}^c$ and $w_{l,k}^d$ are non-zero.
However, it can be shown that, if $R_k$ in the constraint (\ref{BackhaulConstraint_Hybrid}) is fixed, indeed at most one of the two can be non-zero, i.e., a user may only participate in data-sharing or compression, but not both. Intuitively this is due to the fact that if a user's data is shared at a particular BS, it is always better to put all the beamforming power in the data-sharing beamformer, rather than splitting it with the compression beamformer, to avoid the quantization noise penalty associated with the compression process. A more precise statement is given below.

\begin{proposition}
	\label{prop:hybrid}
	The global optimal point $(\{w_{l,k}\},\{w_{l,k}^c\},\{w_{l,k}^d\})$ to the optimization problem (\ref{hybrid_original_optimization}) with fixed $R_k$ in the constraint (\ref{BackhaulConstraint_Hybrid}) has either $w_{l,k}^c = 0$, or $w_{l,k}^d = 0$, or both for all $l \in \mathcal{L}, k \in \mathcal{K}$.
\end{proposition}

The proposition can be proved using contraction by showing that if both
$w_{l,k}^c$ and $w_{l,k}^d$ are non-zero, then we can construct another
feasible solution that strictly improves the weighted sum rate, thus such a solution cannot be optimal.

\subsection{Optimization Methodology}

The problem (\ref{hybrid_original_optimization}) involves a joint optimization of beamforming vector for compression and data-sharing signals ($\{w_{l,k}^c,w_{l,k}^d\}$) (and as a consequence the combined beamformers $\{w_{l,k}\}$), the quantization noise levels $\{q_l\}$ for the compression signal, and the BS clustering for data-sharing (and thus compression), i.e., the decision of which users should data-shared and which users should be compressed for which BSs. 
The problem is challenging as it combines the difficulties in optimizing the
individual data-sharing and compression strategies. This paper proposes a
solution strategy as follows.

 The main source of difficulty is the constraint (\ref{BackhaulConstraint_Hybrid}). The first term is the indicator function accounting for backhaul consumption due to direct data-sharing, along with the user rate $R_k$ that is also part of the objective function. The second term with the $\log$ function in the compression part is a non-convex function of the variables ($\{w_{l,k}^c\},\{q_l\}$). For the indicator function, as before we approximate it as a weighted $l_1$ norm as
\begin{equation}
\mathbbm{1}\left\{ \left \vert w_{l,k}^d \right \vert^2 \right\} = \left\| \left \vert w_{l,k}^d \right\vert^2 \right\|_0 \approx \beta_{l,k}^d \left \vert w_{l,k}^d \right \vert^2 ,
\end{equation}
where $\beta_{l,k}^d$ is a constant weight associated with BS $l$ and user $k$ and is updated iteratively in an outer loop according to
\begin{equation} \label{UpdateBeta_Hybrid}
\beta_{l,k}^d = \frac{1}{\left \vert w_{l,k}^d \right \vert^2 + \tau},
\end{equation}
for some regularization constant $\tau > 0$ and $|w_{l,k}^d|^2$ from the previous iteration. Similarly, $R_k$ in the constraint (\ref{BackhaulConstraint_Hybrid}) is kept fixed from the previous iteration, denoted by $\hat{R}_k$, and is updated in the same outer loop. This simplifies the constraint (\ref{BackhaulConstraint_Hybrid}) to
\begin{equation} \label{Approx_BackhaulConstraint_Hybrid}
\sum_{k=1}^K  \beta_{l,k}^d \hat{R}_k \left \vert w_{l,k}^d \right \vert^2 + \log \left( 1 +  \frac{\Gamma_q\sum_{k=1}^K{\left \vert w_{l,k}^c \right \vert^2}}{q_l} \right) \leq C_l, \quad l \in \mathcal{L}.
\end{equation}
Next, we rewrite the $\log$ function in the above constraint (\ref{Approx_BackhaulConstraint_Hybrid}) into sum of two terms for all $l \in \mathcal{L}$ as follows:
\begin{equation}
\sum_{k=1}^K  \beta_{l,k}^d \hat{R}_k \left \vert  w_{l,k}^d \right \vert^2 + \log \left( \Gamma_q\sum_{k=1}^K{\left \vert w_{l,k}^c \right \vert^2} + {q_l} \right) - \log \left( q_l \right ) \leq C_l.  \label{BackhaulConstraint_ApproximatedWSR_Hybrid}
\end{equation}
Thus, we need to solve the optimization problem (\ref{BackhaulConstraint_Hybrid}) with the constraint (\ref{BackhaulConstraint_Hybrid}) modified as (\ref{BackhaulConstraint_ApproximatedWSR_Hybrid}). In this formulation, in the constraint (\ref{BackhaulConstraint_ApproximatedWSR_Hybrid}), $-\log(q_l)$ is a convex function of $\{q_l\}$, but $\log \left( \Gamma_q\sum_{k=1}^K{\vert w_{l,k}^c \vert^2} + {q_l} \right)$ is a non-convex function of ($\{w_{l,k}\},\{q_l\}$). Additionally the objective function (\ref{hybrid_objective}) is a non-convex function of ($\{w_{l,k}\},\{q_l\}$). In order to solve the optimization problem (\ref{BackhaulConstraint_Hybrid}) with the modified constraint (\ref{BackhaulConstraint_ApproximatedWSR_Hybrid}), we use the iterative successive convex approximation method by linearizing the non-convex part in both the objective and the constraint in an inner loop.
First, we transform the objective into a suitable form, by utilizing the relationship between the achievable rate and the MSE. The MSE for user $k$ is defined as
\begin{equation}
\label{eq:MSE_Hybrid}
e_k = \left \vert u_{k} \right \vert^2 v_k - 2 \text{Re} \{ u_{k}^{H} \mathbf{h}_{k}^H \mathbf{w}_k\} + 1,
\end{equation}
under a receive beamformer $u_k$, where
\begin{equation}
v_k = \Gamma_m \Bigg( \sum_{j \neq k} \left \vert \mathbf{h}_{k}^H \mathbf{w}_j \right \vert^2 + \sigma^2 + \mathbf{h}_k^{H}\mathbf{Q}\mathbf{h}_k \Bigg) + \left \vert \mathbf{h}_{k}^H \mathbf{w}_k \right \vert^2.
\end{equation}
The rate $R_k$ can then be written as
\begin{equation}
\label{TransformedObjective_Hybrid}
R_k = \max_{u_k} \log \left(e_k^{-1}\right).
\end{equation}
Second, to deal with the non-convexity of the $\log$ function in the
transformed objective function (\ref{TransformedObjective_Hybrid}) and the
modified constraint (\ref{BackhaulConstraint_ApproximatedWSR_Hybrid}), we find
the appropriate tight convex upper bounds and successively update them. We make
use of the following result, which is a consequence of 
the inequality $\log y \le y -1$ for any positive $y$, with $y = \frac{x}{x_0}$.
\begin{lemma}
	For any positive $x, x_0 \in \mathbb{R}$, $\log x \le \log x_0 + \frac{1}{x_0} x - 1$, with equality if and only if $x = x_0$.
\end{lemma}

We make successive convex approximations to $\log \left ( e_k \right)$ and $\log \left( \Gamma_q\sum_{k=1}^K{\vert w_{l,k}^c \vert^2} + {q_l} \right)$ as follows.
\begin{equation}
\label{rate_approximation}
\log \left ( e_k \right ) \le - \log \left ( \rho_k \right) + \rho_k e_k - 1,
\end{equation}
where the equality holds if 
\begin{equation}
\label{rho_update_hybrid}
\rho_k = e_k^{-1},
\end{equation}
with $e_k$ as defined in (\ref{eq:MSE_Hybrid}), and (\{$w_{l,k}\},\{u_k$\}) taken from the previous iteration in an iterative manner in the inner loop.
Similarly,
\begin{align}
& \log \left ( \Gamma_q\sum_{k=1}^K{\left \vert w_{l,k}^c \right \vert^2} + {q_l} \right) \nonumber \\ & \qquad \le - \log \left ( \gamma_l \right ) + \gamma_l \left ( \Gamma_q\sum_{k=1}^K{\left \vert w_{l,k}^c \right \vert^2} + {q_l} \right )- 1,
\end{align}
where the equality holds if 
\begin{equation}
\label{eq:update_gamma}
\gamma_l = \left ( \Gamma_q\sum_{k=1}^K{\left \vert w_{l,k}^c \right \vert^2} + {q_l} \right)^{-1},
\end{equation}
with (\{$w_{l,k}^c\},\{q_l$\}) taken from previous iteration in the inner loop.

Note the similarity of the update (\ref{rho_update_hybrid}) to the MSE weight update in Section \ref{sec:data_sharing} (i.e., equation (\ref{eq:rho_DataSharing})) used in the iterative algorithm to solve the equivalent WMMSE problem. The two are in fact related. Another way of looking at the iterative algorithm for the equivalent WMMSE problem is exactly what we have done above for the objective function. We successively upper bound the log function in the rate $R_k$ after writing it as a function of the transmit and receive beamformers as in (\ref{TransformedObjective_Hybrid}), then update the convex upper bound in successive block updates in the transmit and receive beamformers. The MSE weights are the multiplying factors in the convex approximations at each step. Such an approach is referred to as block successive upper bound minimization (BSUM) \cite{Razaviyan_siam2013}.

Thus, in the end, we iteratively solve the following programs with alternating block updates in the inner loop for fixed $\beta_{l,k}$ and $\hat{R}_k$, and then update $\beta_{l,k}$ according to (\ref{UpdateBeta_Hybrid}) and $\hat{R}_k$ as the modified $R_k$ in the outer loop, as discussed when simplifying the original constraint (\ref{BackhaulConstraint_Hybrid}) to (\ref{Approx_BackhaulConstraint_Hybrid}). Since the outer loop involves estimates of the user rates from the previous iteration, it is difficult to prove the convergence of the entire algorithm. However, for fixed $\beta_{l,k}$ and $\hat{R}_k$, the inner iterations can be shown to converge due to the convergence properties of BSUM. Moreover, outer iterations are observed to converge in numerical experiments.

\begin{itemize}
	\item The optimal receive beamformer $u_k$ under fixed $\{\mathbf{w}_k,\mathbf{w}_k^c,\mathbf{w}_k^d\}$ and $\{q_l\}$ is given by
	\begin{equation}
	u_k = v_k^{-1} \mathbf{h}_{k}^H \mathbf{w}_k.
	\label{eq:MSEreceiver_Hybrid}
	\end{equation}
	\item Under fixed $\{ u_k \}$, the optimal transmit beamformers $\{\mathbf{w}_k,\mathbf{w}_k^c,\mathbf{w}_k^d\}$ and the optimal quantization noise levels $\{q_l\}$ are obtained by solving the following convex program:
	\begin{subequations} \label{hybrid_approximated_optimization}
		\begin{align}
		\mini_{\substack{\{w_{l,k}^d\},\{w_{l,k}^c\},\\ \{w_{l,k}\},\{q_l\}}} \quad & \sum_{k=1}^K - \alpha_k \rho_k e_k  \\
		\sbto \quad &  \sum_{k=1}^K \left \vert w_{l,k} \right \vert^{2} + q_l \leq P_l, \quad l \in \mathcal{L} \label{PowerConstraint_Hybrid} \\
		& \sum_{k=1}^K \beta_{l,k}^c \hat{R}_k \left \vert w_{l,k}^d \right \vert^{2} + \gamma_{l} \Gamma_q \sum_{k=1}^K \left \vert w_{l,k}^c \right \vert^{2} \nonumber \\ & \quad + \gamma_l q_l- \log(q_l) \leq C'_l, \quad l \in \mathcal{L} \label{hybrid_approximated_optimization_backhaul_constraint}  \\
		& w_{l,k}^d + w_{l,k}^c = w_{l,k}, \quad l \in \mathcal{L}, k \in \mathcal{K}, \label{SumConstraint_Hybrid}
		\end{align}
		where $C'_l = C_l + \log \left ( \gamma_l \right) + 1$.
	\end{subequations}
\end{itemize}

The overall algorithm for the joint optimization of the problem
(\ref{hybrid_original_optimization}) for the hybrid strategy is summarized in
Algorithm \ref{alg:UnifiedHybrid}. Following a similar analysis to the
data-sharing strategy, the computational complexity of the algorithm can be
shown to scale as $O((KLM)^{3.5}T_1T_2)$, where $T_1$ and $T_2$ are the total
number of inner and outer iterations needed to converge, respectively.
\begin{algorithm}[t]
	{\bf Initialization}: $\{\mathbf{w}_k,\mathbf{w}_k^c,\mathbf{w}_k^d\}, \{q_l\}, \{\beta_{l,k}^d\}, \{\hat{R}_k\}$;\\
	{\bf Repeat}:
	\begin{enumerate}
		\item {\bf Repeat}:
		\begin{enumerate}
			\item For fixed $\{\mathbf{w}_k,\mathbf{w}_k^c,\mathbf{w}_k^d\}, \{q_l\}$, compute the optimal receivers $\{u_{k}\}$ according to (\ref{eq:MSEreceiver_Hybrid}) and the corresponding MSE $\{e_k\}$ according to (\ref{eq:MSE_Hybrid});
			\item Update the weights $\{\rho_k\}$ according to (\ref{rho_update_hybrid});
			\item Update the weights \{$\gamma_l$\} according (\ref{eq:update_gamma}), for fixed $\{\mathbf{w}_k^c\}, \{q_l\}$;
			\item For fixed $\{u_{k}\}$, $\{\rho_k\}$, and $\{ \hat{R}_k \}$ in (\ref{hybrid_approximated_optimization_backhaul_constraint}), find the optimal transmit beamformers $\{\mathbf{w}_k,\mathbf{w}_k^c,\mathbf{w}_k^d\}$ by solving (\ref{hybrid_approximated_optimization});
		\end{enumerate}
		{\bf Until} convergence
		\item Update $\{\beta_{l,k}\}$ as in (\ref{UpdateBeta_Hybrid});
		\item Compute the achievable rates $\{ R_k \}$ according to (\ref{eq:Rk_Hybrid}). Update $\hat{R}_k = R_K$, $k \in \mathcal{K}$.
	\end{enumerate}
	{\bf Until} convergence
	\caption{Weighted sum rate maximization for the hybrid strategy}
	\label{alg:UnifiedHybrid}
\end{algorithm}

\section{Numerical Evaluation of the Hybrid Strategy}
\label{sec:hybrid_evaluation}

\begin{figure}[t]
	\centering
	\includegraphics[trim=0.75cm 0.05cm 0.75cm 0.10cm, clip,
	width=1.0\columnwidth]{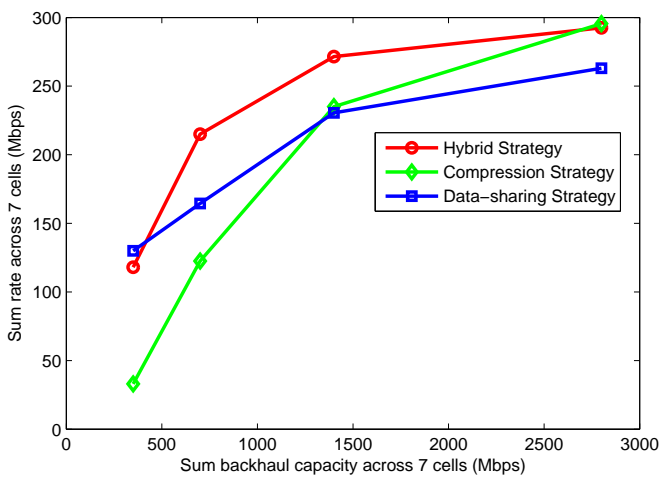}
	\caption{Comparison of the hybrid strategy with the data-sharing and compression strategies.}
	\label{fig:comparison_sumrateVSsumbackhaul_explicit}
\end{figure}

We evaluate the hybrid strategy for the 7-cell wrapped around two-tier heterogeneous
network considered in Section \ref{sec:data_sharing_vs_compression}.
All the BSs and the user terminals are equipped with a single antenna each.
We compare the hybrid strategy designed with the joint optimization done by
Algorithm \ref{alg:UnifiedHybrid}, with the data-sharing and
compression strategies, optimized with explicit per-BS backhaul
constraints using Algorithm \ref{alg:LCSparse_MIMO} and Algorithm
\ref{alg:pureCompression}, respectively.

Fig.~\ref{fig:comparison_sumrateVSsumbackhaul_explicit} shows the average sum rate as a function of total backhaul capacities across the 7-cell network for the three strategies. For low backhaul capacity, we observe that the data-sharing performs better than the compression strategy. In this case, the hybrid strategy performs just as well, because almost all the users in the final beamformer for the hybrid strategy are data-shared. The reason that the hybrid strategy is seen to be slightly worse than the data-sharing strategy at very low backhaul capacity is due to numerical precision. For moderate backhaul capacity, the data-sharing and compression strategies are comparable. This is the regime where hybrid strategy has some potential to provide benefits by having some users participate in data-sharing and rest in the compression. When the backhaul capacity is high, the compression strategy starts to outperform the data-sharing strategy. The hybrid strategy shows some improvement in this regime and the gains tend to diminish as we increase the backhaul even further as the rates saturate to the maximum sum rate of the system. Thus, overall we see that the hybrid strategy achieves the best of the two strategies under low and high backhaul capacities, and when the backhaul capacities are moderate, there is some benefit from the hybrid design.

\section{Conclusions}
\label{sec:conclusions}

This paper compares two fundamentally different strategies, the compression
strategy, which is the standard solution for C-RAN, and the data-sharing
strategy, which is the traditional implementation in most cellular systems. Our
main conclusion is that fronthaul/backhaul capacity constraint is crucial in
deciding which strategy should be adopted for the downlink C-RAN. If the
available fronthaul/backhaul capacity is medium-to-high, the compression
strategy outperforms the data-sharing strategy due to the possibility of having
large cooperation cluster at the central processor, whereas using data-sharing,
the cluster size is limited by the backhaul capacity. However, if the available
fronthaul/backhaul capacity is low, the data-sharing strategy outperforms the
compression strategy. Under low fronthaul capacity, the quantization noises
introduced in the compression strategy dominate the interference, in which case
it is better to share the data directly with a limited set of BSs rather than
to compress.  When we also take into account the cost of CSI acquisition, the
performance of the compression strategy suffers more from the lack of CSI than
data-sharing. This is because the gain in the compression strategy
stems from the possibility of forming larger cooperation clusters at the central
processor, which requires more CSI than the data-sharing strategy, which already 
has a smaller cluster size due to the limited backhaul capacity.

Motivated by such comparison, this paper proposes to combine the data-sharing
and compression strategies into a hybrid scheme that can benefit from the
advantages of both strategies. Such hybrid combination results in flexibility
in terms of backhaul utilization. The optimization framework proposed for the
hybrid strategy generalizes both individual strategies. When the backhaul
capacity is low, the hybrid strategy reduces to primarily data-sharing and when
the backhaul capacity is high, it reduces to almost all compression. But when
the fronthaul/backhaul capacity is moderate, we observe that the system
performance can be improved by sharing the data for some of the users directly
with the BSs and sending compressed version of the signals of the rest of the
users using the remaining fronthaul/backhaul capacity.  Having the flexibility
to switch between data-sharing and compression depending on the available
fronthaul/backhaul capacity at different BSs is especially useful in the future
dense cellular networks with different tiers of BSs, with different levels of
fronthaul/backhaul capacities.

\bibliographystyle{IEEEtran}
\bibliography{IEEEabrv,references}

\end{document}